\documentclass[paper]{jhep3}
\usepackage{epsfig,multicol}
\usepackage{cite}
\usepackage{epsfig}
\usepackage{amsmath}
\usepackage{amssymb}

\preprint{KEK-TH-1067 \\ KYUSHU-HET-91}

\title{Dynamical aspects of
the plane-wave matrix model\\
at finite temperature}

\author{ 
Naoyuki Kawahara${}^{ab}$\,, Jun Nishimura${}^{ac}$ 
and Kentaroh Yoshida${}^{a}$ 
\vspace*{0.5cm} \\ 
\llap{$^a$}High Energy Accelerator Research Organization (KEK),\\
1-1 Oho, Tsukuba 305-0801, Japan  \\
\llap{$^b$}Department of Physics, Kyushu University, 
Fukuoka 812-8581, Japan \\ 
\llap{$^c$}Department of Particle and Nuclear Physics,\\
Graduate University for Advanced Studies (SOKENDAI),\\
1-1 Oho, Tsukuba 305-0801, Japan 
\vspace*{0.5cm} \\ 
\email{kawahara@post.kek.jp,
jnishi@post.kek.jp, kyoshida@post.kek.jp}
}

\abstract{
We study dynamical aspects of 
the plane-wave matrix model at finite temperature.
One-loop calculation around general classical vacua
is performed using the background field method,
and the integration over the gauge field moduli is carried out
both analytically and numerically.
In addition to the trivial vacuum, 
which corresponds to a single M5-brane at zero temperature,
we consider general static fuzzy-sphere type configurations.
They are all 1/2 BPS, and hence degenerate
at zero temperature due to supersymmetry.
This degeneracy is resolved, however, at finite temperature,
and we identify the configuration that gives the smallest free energy
at each temperature.
The Hagedorn transition in each vacuum is
studied by using the eigenvalue density method 
for the gauge field moduli,
and the free energy as well as the Polyakov line 
is obtained analytically near the critical point.
This reveals the existence of fuzzy sphere phases, 
which may correspond to the plasma-ball phases 
in $\mathcal{N}=4$ $SU(\infty)$ SYM on $S^1 \times S^3$.
We also perform Monte Carlo simulation to integrate
over the gauge field moduli.
While this confirms the validity of the analytic 
results near the critical point, it also shows
that the trivial vacuum gives the smallest free energy
throughout the high temperature regime.
}


\def\be{\begin{equation}}
\def\ee{\end{equation}}
\def\beq{\begin{equation}}
\def\eeq{\end{equation}}
\def\bear{\begin{eqnarray}}
\def\eear{\end{eqnarray}}
\def\beqa{\begin{eqnarray}}
\def\eeqa{\end{eqnarray}}
\def\nn{\nonumber}

\newcommand{\dt}{\partial_t}
\newcommand{\Dt}{D_t}
\newcommand{\bDt}{\overline{D}_t}
\newcommand{\half}{\frac{1}{2}}
\newcommand{\tri}{\frac{1}{3}}
\newcommand{\quarter}{\frac{1}{4}}

\newcommand{\mt}{\frac{1}{3}}
\newcommand{\ms}{\frac{1}{6}}
\newcommand{\mf}{\frac{1}{4}}

\newcommand{\mts}{\frac{1}{9}}

\newcommand{\ijk}{\epsilon_{ijk}}
\newcommand{\bA}{\bar{A}}
\newcommand{\tA}{\tilde{A}}
\newcommand {\tr}{{\rm tr\,}}

\begin{document}

\section{Introduction}

Matrix models provide a promising approach to non-perturbative
dynamics of superstring theories and M-theory \cite{BFSS,IKKT,DVV}. 
In particular the matrix model proposed by Banks,
Fischler, Shenker and Susskind (BFSS) \cite{BFSS} is conjectured 
to be a non-perturbative formulation of M-theory, 
a hypothetical theory in eleven dimensions,
whose low energy effective theory is given by 11d supergravity. 
The BFSS matrix model, which takes the form of a
matrix quantum mechanics, can be obtained also from the 
supermembrane theory in eleven dimensions via
the matrix regularization \cite{dWHN}. 
Studying the BFSS matrix model at finite temperature is
an interesting subject due to its relation to the 
Schwarzschild black hole \cite{KS,BFKS,KLL}.
Such studies have been performed, for instance, in
refs.\ \cite{AMS,others}. 
Applying matrix models to cosmology is
another fascinating direction, 
which has developed recently \cite{Mat-cos}.

While the BFSS matrix model is defined on a flat Minkowski space-time,
we may consider matrix models also on curved space-time.
In this regard the so-called pp-wave background provides
a particularly simple and tractable example.
In fact the maximally supersymmetric pp-wave background
in eleven dimensions, which preserves 32 supercharges, 
is unique and its explicit form is known
\cite{KG}. 
Berenstein, Maldacena and Nastase \cite{BMN}
proposed a matrix model on this background,
which is closely related to the supermembrane theory on the pp-wave 
through the matrix regularization \cite{DJR,SY,NSY}.
It has fuzzy spheres as classical solutions
due to the presence of the mass term and the Myers term \cite{Myers}. 
The stability of these solutions \cite{DJR,SY3,HSKY},
as well as the spectrum around them 
\cite{DJR,DJR2,KP,KP2}, has been studied intensively,
while more general classical solutions such as a rotating
fuzzy sphere are also discussed
\cite{Bak,HS1,Park,sol,HSKY,HSKY-potential} \footnote{A matrix string
theory on a type IIA plane-wave background, which includes fuzzy spheres
as classical solutions, has originally been constructed in \cite{SY4}. 
The spectrum around the fuzzy spheres was computed in \cite{DMS} by
following the method of \cite{DJR}. The theory is applicable to the
matrix cosmology scenario \cite{DM}.}.  
In fact the fuzzy sphere solutions can be interpreted as giant gravitons,
and the interaction potential between them
\footnote{The interaction potential between point-like gravitons
has been studied recently in ref.\ \cite{SY-point}. 
See also ref.\ \cite{SY-review}
for a short review on the graviton potential.
}
is shown to be of the $1/r^7$-type \cite{HSKY,HSKY-potential}
similarly to the results in the BFSS matrix model. 
This type of potential is anticipated since
the spectrum in the linearized eleven-dimensional supergravity around
the pp-wave background \cite{Kimura} is included in the matrix model as
the spectrum of the zero-mode Hamiltonian \cite{KP,NSY}.


The plane-wave matrix model has been studied 
also at finite temperature by various authors.
In ref.\ \cite{FSS} the free energy around the trivial vacuum,
which corresponds to a transverse M5-brane \cite{TM5} at
zero temperature \footnote{In a deformed plane-wave matrix model with an
interaction term coming from the 6-form potential, a fuzzy five-sphere
solution was constructed \cite{YL}.},
was evaluated at the one-loop level, and the Hagedorn transition
was studied in detail.
(See refs.\ \cite{SRV,Semenoff2} for a two-loop extension
and ref.\ \cite{Semenoff} for a review on this subject.) 

In this paper we extend such calculations to general classical vacua,
and identify the vacuum that gives the smallest free energy
at each temperature.
Similar studies
have been made earlier in refs. \cite{Huang, HSKY-thermal}.
However, the degrees of freedom corresponding to 
the gauge field moduli, which appear at finite temperature,
were not taken into account
although they actually play an important role 
in the Hagedorn transition \cite{FSS}.
Here we perform the one-loop calculation
by the background field method keeping the gauge field moduli arbitrary.
The integration over the gauge field moduli is performed 
both analytically and numerically.
Near the critical point
we use the eigenvalue density method 
to obtain analytical results
extending the previous works \cite{FSS,Semenoff} 
for the trivial vacuum.
This reveals the existence of
fuzzy-sphere phases, 
which may correspond to the plasma-ball phases in 
$\mathcal{N}=4$ SU($\infty$) SYM on $S^1 \times S^3$. 
We also perform
Monte Carlo simulation 
to integrate over the gauge field moduli
with the matrix size up to $N =2000$, which allows
a reliable extrapolation to $N =\infty$.
(Note that the results of ref.\ 
\cite{Huang, HSKY-thermal} are obtained for small $N$.)
While this confirms the validity of our analytical results,
it also shows that the trivial vacuum gives the smallest
free energy throughout the high temperature regime.
In fact the 
leading asymptotic behavior in the high temperature limit
is universal, but the difference appears in the subleading term.


Comparison of the free energy for general classical vacua was 
also made in a matrix model,
which is obtained by dimensionally reducing
the plane-wave matrix model to a point \cite{massiveIIB}.
There the true vacuum turned out to be a complicated multi-fuzzy-sphere
type configuration, which
incorporates a non-trivial gauge group.
\footnote{
Note that the model studied there
is closely related, but not equivalent 
to the high temperature limit of the plane-wave
matrix model.
}
Analogous models without the mass term have been studied intensively
by Monte Carlo simulation \cite{ABNN}.
These studies are partly motivated from the IIB matrix model \cite{IKKT},
a conjectured nonperturbative formulation 
of type IIB superstring theory,
which can be formally obtained by dimensionally reducing
the BFSS matrix model to a point. 
More intimate relationship between the two models
has been suggested \cite{Kawahara:2005an} on account of
the large $N$ reduction \cite{EK}. 
Calculation of the free energy in the IIB matrix model
for various configurations provided certain evidences that
four-dimensional space-time is favored dynamically \cite{SSB}.
See ref.\ \cite{related} for related works on this issue.

The results obtained in the present paper
are expected to have close relationship to 
the phase structure of large $N$ gauge theories at finite temperature
and at finite volume, which has been explored in refs.\ 
\cite{Sundborg,Aharony1,Aharony2,Aharony:2005cx,SV}.
According to the AdS/CFT correspondence 
\cite{AdS} at finite temperature
\cite{Witten:thermal},
the observed phase transitions should correspond to
those on the gravity side
such as the Hawking-Page phase transition \cite{HP}
and the Gregory-Laflamme phase transition \cite{GL}.
This provides an explicit realization of the connection 
between the Hagedorn transition in string theory 
and the deconfinement phase transition in large $N$ gauge theories 
pointed out earlier \cite{AW}.

The rest of this paper is organized as follows.
In section \ref{section:classical}
we briefly review the plane-wave matrix model and 
its classical vacua.
In section \ref{section:one-loop}
we perform one-loop calculation
around general classical vacua
keeping the gauge field moduli arbitrary.
In section \ref{section:thermo-pp}
we perform the integration over the gauge field moduli
to obtain explicit results for the free energy and the
Polyakov line.
Section \ref{section:summary}
is devoted to a summary and discussions. 
In the Appendices we present the details of our calculations.

\section{The plane-wave matrix model and its classical solutions}
\label{section:classical}

In this section we define the plane-wave matrix model
and discuss its classical solutions.
In order to study the model at finite temperature,
we make the Wick rotation $t \rightarrow -it$
and compactify the imaginary time direction
to a circle with the circumference
$\beta\equiv 1/T$. Thus the action can be written as
\footnote
{We have made the rescaling, 
$A \rightarrow RA\,,~t \rightarrow \frac{1}{R} t\,,
~\mu \rightarrow R \mu\,,$ in the action presented, for instance,
in ref.\ \cite{SY} to arrive at the present form,
which does not include the parameter $R$
(the compactification radius of the 11th direction in M-theory)
explicitly.}
\begin{eqnarray}
S_{\rm pp} &=& \int_0^{\beta}\!\! dt\, {\rm Tr} 
	\left[
	\half (\Dt X_M)^2 - \quarter [X_M,X_N]^2 
	+ \half \Psi^{\dagger}\Dt\Psi - \half \Psi^{\dagger}\gamma_M[X_M,\Psi]
	\right. \nn \\
&{}& \left.
	+ \half \left(\frac{\mu}{3}\right)^2
  (X_i)^2 + \half \left(\frac{\mu}{6}\right)^2  (X_a)^2 
	+ i \frac{\mu}{3}  \epsilon_{ijk} X_i X_j X_k 
 + i \frac{\mu}{8}  \Psi^{\dagger} \gamma_{123} \Psi
	\right] \, ,
\label{BMN-action}
\end{eqnarray}
where $\Dt = \dt - i[A, \ \cdot \ ]$\, represents
the covariant derivative, and the indices $M,N$ run from 1 to 9.
Since the transverse $SO(9)$ symmetry is broken down to 
$SO(3)\times SO(6)$ on the pp-wave background,
we have also introduced the $SO(3)$ indices
$i,j,k = 1,2,3$ and the $SO(6)$ indices $a = 4,\cdots , 9$\,. 
The partition function for the finite temperature system is defined by
\begin{eqnarray}
Z &=& \int [dA(t)][dX_M(t)][d\Psi(t)][d\Psi^{\dagger}(t)] 
	\exp\left(-S_{\rm pp}\right) \ , 
\end{eqnarray}
where the bosonic and fermionic fields obey the periodic and 
anti-periodic boundary conditions, respectively, 
in the imaginary time direction. 

Let us discuss static classical solutions of the model
within the usual Ansatz $X_a = \Psi =0$\,.
At zero temperature it is convenient to set the 
one-dimensional gauge field $A(t)$ to zero 
by choosing the gauge, while at finite temperature 
we can only set $A(t)$ to a constant diagonal matrix
due to the nontrivial holonomy in the temporal direction.
In the case of the trivial vacuum $X_i=0~(i=1,2,3)$,
for instance, the gauge field can be an arbitrary constant diagonal
matrix without increasing the action, 
and the corresponding moduli parameters should be integrated.
These degrees of freedom,
which we refer to as the gauge field moduli,
indeed play a crucial role in the Hagedorn transition \cite{FSS}.

In fact the plane-wave matrix model has static fuzzy-sphere type
classical solutions
\begin{eqnarray}
\label{fuzzy-sphere0} 
X_i &=& B_i  \equiv
\frac{\mu}{3}L_i
= \frac{\mu}{3} \bigoplus_{I=1}^{s}
\Bigl( {L_i^{(n_I)}} \otimes {\bf 1}_{k_I} \Bigr)
\, , \\
A(t) &=& \bar{A}(t) 
\equiv 
\bigoplus_{I=1}^{s}
\Bigl( 
{\bf 1}_{n_I} \otimes \bar{A}^{(I)}(t)
\Bigr) \ ,
\label{fuzzy-sphere1} 
\end{eqnarray} 
where $L_i^{(n)}$ are the $SU(2)$ generators 
in the $n$-dimensional irreducible representation
satisfying $[L_i^{(n)},L_j^{(n)}] = i\epsilon_{ijk} L_k^{(n)}\, ,$
and the parameters $k_I$ and $n_I$ obey
$
\sum_{I=1}^{s}n_I \cdot k_I \nn  = N \ .
$
The $k_I \times k_I$ matrices
$\bar{A}^{(I)}(t)$ represent the gauge field moduli.
One can easily check that the classical action for the above
solution is zero for arbitrary $\bar{A}^{(I)}(t)$
by noting that $[B^i,\bar{A}(t)]=0$\,.

At zero temperature, 
these solutions are interpreted as stacks of M2-branes or
M5-branes (or their mixture) depending on how one takes the
large $N$ limit. In what follows we often restrict ourselves
to the $s=1$ case for simplicity, and use the parameters
$n\equiv n_1$ and $k\equiv k_1$, which satisfy $n\cdot k = N$.
Then the solution represents $k$ copies of spherical M2-branes
if $k$ is held fixed, while it represents
$n$ copies of spherical M5-branes if $n$ is held fixed,
in the large $N$ limit.
An evidence for the latter interpretation
is provided by the coincidence of 
the spectrum \cite{TM5} for the trivial vacuum ($n=1$) case.
Note, however, that 
the argument for identifying transverse M5-branes in the
plane-wave matrix model is based on the protected BPS multiplet,
and it relies rather crucially on supersymmetry,
which is broken at finite temperature. 
Hence it is not totally clear whether the M5-branes
are stable against thermal fluctuations. 
One way to demonstrate the existence of M5-branes 
is to compute their radius at finite temperature 
by the Gaussian expansion method extending the calculation in
ref.\ \cite{KLL} to the present case.

\section{One-loop effective action}
\label{section:one-loop}

In this section we compute the one-loop effective action 
around the classical solutions keeping the gauge field moduli
arbitrary using the background field method. 
The matrices are decomposed 
as
\begin{eqnarray}
&& X_i(t) = B_i + Y_i(t)\ , \qquad 
X_a(t) = 0 + Y_a(t)\ ,  \nn \\
&&  A(t) = \bar{A}(t)+\tilde{A}(t)\ , \qquad 
\Psi(t) = 0+\Psi(t)\ ,
\label{background+fluctuation}
\end{eqnarray} 
where $Y_i(t)$, $Y_a(t)$, $\tilde{A}(t)$ and $\Psi(t)$ represent
the fluctuations around the background $B_i$ and $\bar{A}(t)$.
 
For the trivial vacuum $X_i=0$\,, 
since the fluctuation $\tA(t)$ can be totally absorbed into the gauge
field moduli $\bA(t)$, we set $\tA(t)$ to zero.
Similarly, in the general case, 
we omit
$\sum_{I=1}^{s} (k_I)^2$ zero modes
in the fluctuation $\tA(t)$,
which corresponds to changing the gauge field moduli.
In the presence of the background, 
the $U(N)$ gauge symmetry 
is broken down to $\prod_{I=1}^{s}U(k_I)$.
Therefore, we also obtain zero modes 
in the direction of the gauge orbit
corresponding to the broken symmetry.
These zero modes should be removed by an appropriate gauge fixing,
and we use the background field gauge for this purpose.
By using the unbroken gauge symmetry,
the gauge field moduli 
$\bar{A}^{(I)}(t)$ can be brought 
into a static diagonal form 
\begin{eqnarray} 
\bA^{(I)}(t) = \frac{1}{\beta}
{\rm diag}(\alpha^{(I)}_1 , \cdots  , \alpha^{(I)}_{k_I}) \, ,
\label{gauge-fixing3} 
\end{eqnarray}
%
where $\alpha_{a}\in (-\pi , \pi]$\,. 
See Appendix \ref{section:gauge-fixing}
for the details of the two gauge fixing procedures.

In the rest of this section, we integrate out
the fluctuations perturbatively keeping the moduli parameters
$\alpha^{(I)}_{a}$ ($a = 1 , \cdots , k_I$) arbitrary.
We add the gauge fixing term $S_{\rm g.f.}$
and the ghost term $S_{\rm ghost}$ 
given by eq.\ (\ref{gf-ghost-terms})
to the action, and expand it with respect to the fluctuations
as $S = S^{(0)}+ S^{(1)}+ \cdots + S^{(4)}$.
The classical action $S^{(0)}$ vanishes for the backgrounds 
we consider here,
and the linear term $S^{(1)}$ vanishes, too,
since the backgrounds satisfy the classical equations of motion.
The quadratic term is given by
\begin{eqnarray}
S^{(2)} &=& \int_0^{\beta}\!\!dt\, {\rm Tr}  
	\left[
	\half (\bDt Y_i)^2 - [Y_i,Y_j][B_i,B_j] - \half [Y_i,B_j]^2 
	+ \half \left(\frac{\mu}{3}\right)^2
 Y_i^2 \right. \nn \\ 
&& \qquad\qquad - i \frac{\mu}{2}
 \ijk Y_i \left[B_k,Y_j\right] 
+ \half (\bDt Y_a)^2 - \half [Y_a,B_i]^2 + 
\half \left(\frac{\mu}{6}\right)^2  Y_a^2 \nn \\ 
&& \qquad\qquad  + \psi^{\dagger A \alpha} \bDt \psi_{A \alpha} 
   + \psi^{\dagger A \alpha} \sigma^{\beta}_{i\alpha}[B_i,\psi_{A \beta}] 
   + \frac{\mu}{4}
 \psi^{\dagger A \alpha}\psi_{A \alpha} \nn \\
&& \qquad\qquad  \left.
+ \half (\bDt \tA)^2 - \half [\tA,B_i]^2 
	+ \bDt \bar{c} \cdot \bDt c - [B_i,\bar{c}][B_i,{c}]
	\right]\, , 
\label{1-loop1} 
\end{eqnarray}
where $c$ and $\bar{c}$ are the Faddeev-Popov ghosts.
We have decomposed the spinor indices according 
to $SU(2) \times SU(4)$,
where $\alpha=1,2$ and $A = 1,\cdots,4$ represent
$SU(2)$ and $SU(4)$ indices, respectively \cite{DJR}. 
The contributions of the higher order terms can be neglected
in the $\mu\to\infty$ limit, 
as can be seen by rescaling of the variables as
\begin{eqnarray}
&& Y_M \to \mu^{-1/2}Y_M 
\ , \quad \tilde{A} \to \mu^{-1/2}\tilde{A}\ , \quad 
\bar{A} \to \mu \bar{A}\ , \nn\\ 
&& c \to \mu^{-1/2}c\ , \quad \bar{c} \to \mu^{-1/2}\bar{c}\ , \quad 
t \to \mu^{-1} t\ , 
\label{rescaling-mu}
\end{eqnarray} 
which brings the action $S$ into the form
\[
S = S^{(2)} + \mu^{-3/2}S^{(3)} + \mu^{-3} S^{(4)}\ ,
\]
where $\mu$ that appear
in $S^{(2)}$\,, $S^{(3)}$ and $S^{(4)}$ are set to unity.
Hence the one-loop calculation is justified in the large $\mu$ limit.
Since we also take the large $N$ limit, it is important
to clarify the $N$ dependence of the expansion parameter.
Let us recall that the expansion parameter in
perturbation theory is given by 
$ \frac{NR^3}{\mu^3} =
\frac{N^4}{(\mu p^+)^3}\,, $
where we have temporarily restored the parameter $R$. 
Clearly the expansion parameter diverges in the large
$N$ limit with fixed $\mu$ and $p^+=N/R$. 
Hence, in order to ensure the validity of the perturbative expansion,
we need to take the $\mu\to\infty$ limit faster than the large $N$ limit.  
Note also that non-perturbative effects such as 
the tunneling between the trivial vacuum and the fuzzy sphere vacua is
suppressed due to the $N\to \infty$ and $\mu\to\infty$ limits \cite{YY}.
With this understanding, we set $\mu = 1$ to simplify the expressions
in what follows, but one can restore the $\mu$ dependence of 
the results after integrating the fluctuations by simply replacing
$\beta$ by $\beta \mu$ (i.e., replacing $T$ by $\frac{T}{\mu}$).



Let us introduce the operators
\begin{eqnarray}
 \mathcal{L}_i M   \equiv [L_i,M] \ ,  
\end{eqnarray}
which act on an $N \times N$ matrix $M$.
Then the quadratic term $S^{(2)}$ in the action can be rewritten as 
\begin{eqnarray}
\label{quadS-rewritten}
&& S^{(2)}  = \int_0^{\beta}\!\! dt\, {\rm Tr} 
	\left[
	\half (\bDt Y_i)^2 + \half Y_i P_{ij}(\mathcal{L}) Y_j 
	+ \half (\bDt Y_a)^2 + \half Y_a Q(\mathcal{L}) Y_a
	\right. \label{1-loop2} 
 \\ &{}& \quad  \left. 
+ \psi^{\dagger A \alpha} \bDt \psi_{A \alpha}  
	+ \psi^{\dagger A \alpha} R^{\beta}_{\alpha}(\mathcal{L})
          \psi_{A \alpha} 
	+ \half (\bDt \tA)^2 + \half \tA T(\mathcal{L}) \tA
	+ \bDt \bar{c} \bDt c + \bar{c} T(\mathcal{L}) c 
	\right]\, , \nn
\end{eqnarray}
where we have defined the following mass operators
\begin{eqnarray} 
&& P_{ij}(\mathcal{L}) 
	= \mts \left\{ \left( \mathcal{L}_k{}^2+1 \right) \delta_{ij}
	- i \ijk \mathcal{L}_k \right\}\ , \qquad  
Q (\mathcal{L}) 
	= \mts \left( \mathcal{L}_k{}^2 + \quarter \right) \ , \nn \\ 
&& R^{\beta}_{\alpha} (\mathcal{L}) 
	=  \quarter \delta^{\beta}_{\alpha} 
	+ \tri \sigma^{\beta}_{k \alpha} \mathcal{L}_k \ , \qquad 
T (\mathcal{L}) = \mts \mathcal{L}_k{}^2 \ . 
\label{mass-operator}
\end{eqnarray} 
The mass spectra, given by the eigenvalues of these operators,
have been obtained for the physical modes \cite{DJR} and
for the unphysical modes \cite{HSKY}.
Due to the structure (\ref{fuzzy-sphere0}) of $L_i$, 
it suffices to solve the eigenvalue problem for
the ($n_I \times n_I$) square block and
the ($n_I \times n_J$) rectangular block
in the matrix, on which the mass operators act.
The results are summarized in
tables \ref{spectrum1} and \ref{spectrum2}, respectively.
Note that the mass spectra for the square block 
in the gauge and ghost fluctuations
do not contain massless modes unlike the spectra
obtained in ref.\ \cite{HSKY} for the vanishing background gauge field.
In fact those massless modes are treated exactly as
the gauge field moduli and an appropriate gauge fixing
in the present formulation.

\TABLE[ht]{\footnotesize 
\begin{tabular}{|c|c|c|c|}	\hline
	type of fluctuations & mass & spins & degeneracy \\ \hline\hline
	$Y_i$ (i=1,2,3) & $\mt \sqrt{l(l+1)}$ & $1 \leq l \leq n_I-1$
			 & $2l+1$ \\ 
	      & $\mt (l+1)$ & $0 \leq l \leq n_I-2$ 
			 & $2l+1$ \\
	      & $\mt l$ & $1 \leq l \leq n_I$ 
			& $2l+1$ \\ \hline
	$Y_a(a=4,\cdots,9)$ & $\mt (l+\half)$ & $0 \leq l \leq n_I-1$ 
			 & $6(2l+1)$ \\ \hline
	$\psi$ (fermion) & $(\frac{l}{3}+\quarter)$ 
			& $\half \leq l \leq n_I-\frac{3}{2}$ 
			 & $4(2l+1)$ \\ 
	  	   & $(\frac{l}{3}+\frac{1}{12})$ 
			& $\half \leq l \leq n_I-\half$ 
			 & $4(2l+1)$ \\ \hline
     $\tA$ (gauge) &  $\mt \sqrt{l(l+1)}$ 
			& $1 \leq l \leq n_I-1$ 
			 & $2l+1$ \\ \hline
     $c,\bar{c}$ (ghost) &  $\mt \sqrt{l(l+1)}$ 
			& $1 \leq l \leq n_I-1$ 
			 & $2l+1$ \\ \hline
\end{tabular}
\caption{Mass spectrum for the ($n_I \times n_I$) square block.}
\label{spectrum1}
}

\TABLE[ht]{\footnotesize 
\begin{tabular}{|c|c|c|c|}	\hline
	type of fluctuations & mass & spins  & degeneracy \\ \hline\hline
	$Y_i(i=1,2,3)$ & $\mt \sqrt{l(l+1)}$ & $\frac{1}{2}|n_I-n_J| 
\leq l \leq \frac{1}{2}(n_I+n_J)-1$
			 & $2l+1$ \\ 
	      & $\mt (l+1)$ & $\frac{1}{2}|n_I-n_J|-1 \leq l \leq 
\frac{1}{2}(n_I+n_J)-2$ 
			 & $2l+1$ \\
	      & $\mt l$ & $\frac{1}{2}|n_I-n_J|+1 \leq l \leq 
\frac{1}{2}(n_I+n_J)$ 
			 & $2l+1$ \\ \hline
	$Y_a(a=4,\cdots,9)$ & $\mt (l+\half)$ & $\frac{1}{2}|n_I-n_J| \leq l 
\leq \frac{1}{2}(n_I+n_J)-1$ 
			 & $6(2l+1)$ \\ \hline
	$\psi$ (fermion) & $(\frac{l}{3}+\quarter)$ 
& $\frac{1}{2}|n_I-n_J|-\half \leq l \leq \frac{1}{2}(n_I+n_J)-\frac{3}{2}$ 
			 & $4(2l+1)$ \\ 
	   & $(\frac{l}{3}+\frac{1}{12})$ 
	& $\frac{1}{2}|n_I-n_J|+\half \leq l \leq \frac{1}{2}(n_I+n_J)-\half$ 
			 & $4(2l+1)$ \\ \hline
     $\tA$ (gauge) &  $\mt \sqrt{l(l+1)}$ 
		& $\frac{1}{2}|n_I-n_J| \leq l \leq \frac{1}{2}(n_I+n_J)-1$ 
			 & $2l+1$ \\ \hline
     $c,\bar{c}$ (ghost) &  $\mt \sqrt{l(l+1)}$ 
	& $\frac{1}{2}|n_I-n_J| \leq l \leq \frac{1}{2}(n_I+n_J)-1$ 
			 & $2l+1$ \\ \hline
\end{tabular}
\caption{Mass spectrum for the ($n_I \times n_J$) rectangular block.}
\label{spectrum2}
}

Integrating out the fluctuations,
we obtain the determinant of operators 
appearing in the quadratic action (\ref{quadS-rewritten}), 
which can be evaluated in a standard way, once
the mass spectra are given.
(See Appendix \ref{section:eval-det}.)
Including the Vandermonde determinant, which comes from the gauge
fixing for the moduli integration (See
Appendix \ref{section:gf-moduli}.), 
the partition function is obtained at the one-loop level as 
\begin{eqnarray}
Z
\equiv \mathcal{C} 
\int [d \alpha] \exp(-S_{\rm eff}[\alpha])\ ,
\label{partition-fn-oneloop}
\end{eqnarray}
where $S_{\rm eff}[\alpha]$ represents the effective action
for the moduli parameters $\alpha_{a}^{(I)}$,
and the integration measure $[d\alpha]$ is given by 
\[
[d\alpha] = \prod_{I=1}^{s} \frac{1}{k_I!}
\prod_{a=1}^{k_I} \frac{d \alpha_{a}^{(I)}}{4\pi T}\, .
\] 
The normalization constant $\mathcal{C}$ turns out to be 
the same for all the classical solutions, as we demonstrate
in the Appendix \ref{section:overall}, and therefore it is 
irrelevant.

For the trivial vacuum, the masses are given by $\mt$\,,
$\ms$ and $\mf$ for the fluctuations
$Y_i$, $Y_a$ and $\psi$, respectively,
and the one-loop effective action $S_{\rm eff}$ reads \cite{FSS}
\begin{eqnarray}
S_{\rm eff} &=& \sum_{a,b=1}^{N}
\left[3 \ln\sinh\left\{\frac{1}{2} 
\left( \mt \beta + i(\alpha_{a}-\alpha_{b}) \right) \right\} 
+ 6 \ln\sinh\left\{\frac{1}{2}\left(\ms \beta +i(\alpha_{a}-\alpha_{b}) 
\right) \right\} \right. \nn \\	
&{}& \left. - 8 \ln \cosh \left\{ \frac{1}{2} \left( \mf \beta
+i(\alpha_{a}-\alpha_{b}) \right) \right\}\right] - \sum_{a \ne b}^{N} 
\left\{ \ln \left| 
\sin \left(\frac{1}{2} (\alpha_{a}-\alpha_{b})
 \right) \right|\right\}\, .
\label{effective-action1}
\end{eqnarray}
In the general case, the one-loop effective action 
$S_{\rm eff}$ can be decomposed as
\begin{eqnarray}
S_{\rm eff} = \sum_{I=1}^{s} S_{\rm eff}^{(I)} + 
\sum_{I \neq J}^{s} S_{\rm eff}^{(I,J)}\, , 
\label{eff2}
\end{eqnarray}
where the first (second) term comes from integrating
the square (rectangular) block of the fluctuations,
which corresponds to the interaction between the ``fuzzy spheres''
represented by matrices of the same (different) size, respectively.
One series of the spectrum for $Y_i$ gives $\mt\sqrt{l(l+1)}$,
which coincides, including degeneracy, with the spectrum 
for the gauge field fluctuation $\tA$ and ghost fields 
$c$, $\bar{c}$. These are the unphysical modes, 
which cancel each other exactly.
Thus, each term in eq.\ (\ref{eff2}) is given by
\begin{eqnarray}
S_{\rm eff}^{(I)} &=& \sum_{a,b=1}^{k_I} 
	\left[
	 \sum_{l=0}^{n_I-2} 
		(2l+1) \ln \sinh 
\left\{ \frac{1}{2} 
	\left( \mt \beta (l+1) +i(\alpha_{a}^{(I)}-\alpha_{b}^{(I)}) 
			\right)
		\right\} 
	\right. \nn \\
&{}& \left.
	+ \sum_{l=1}^{n_I} 
		(2l+1) \ln \sinh 
		\left\{ \frac{1}{2} 
	\left( \mt \beta l +i(\alpha_{a}^{(I)}-\alpha_{b}^{(I)}) 
			\right)
		\right\} 
	\right. \nn \\
&{}& \left.
	+ \sum_{l=0}^{n_I-1} 
		6(2l+1) \ln \sinh 
		\left\{ \frac{1}{2} 
	\left( \mt \beta (l+\half) +i(\alpha_{a}^{(I)}-\alpha_{b}^{(I)}) 
			\right)
		\right\} 
	\right. \nn \\	
&{}& \left.
	- \sum_{l=\half}^{n_I-\frac{3}{2}} 
		4(2l+1) \ln \cosh 
		\left\{ \frac{1}{2} 
			\left( \beta (\frac{l}{3}+\quarter)
			 +i(\alpha_{a}^{(I)}-\alpha_{b}^{(I)}) 
			\right)
		\right\} 
	\right. \nn \\
&{}& \left.
	- \sum_{l=\half}^{n_I-\half} 
		4(2l+1) \ln \cosh 
		\left\{ \frac{1}{2} 
			\left( \beta(\frac{l}{3}+\frac{1}{12})
			 +i(\alpha_{a}^{(I)}-\alpha_{b}^{(I)}) 
			\right)
		\right\} 
	\right] \nn \\
&{}& 
- \sum_{a \ne b}^{k_I}
\ln \left| \sin 
\frac{\alpha^{(I)}_{a}-\alpha^{(I)}_{b}}{2}
	\right|
\ ,
\label{effective-action2}\\
S_{\rm eff}^{(I,J)} &=& \sum_{a=1}^{k_I}\sum_{b=1}^{k_J} 
	\left[
	 \sum_{l=|n_I-n_J|/2-1}^{(n_I+n_J)/2-2} 
		(2l+1) \ln \sinh 
		\left\{ \frac{1}{2} 
	\left( \mt \beta(l+1) +i(\alpha_{a}^{(I)}-\alpha_{b}^{(J)}) 
			\right)
		\right\} 
	\right. \nn \\
&{}& \left.
	+ \sum_{l=|n_I-n_J|/2+1}^{(n_I+n_J)/2} 
		(2l+1) \ln \sinh 
		\left\{ \frac{1}{2} 
		\left( \mt \beta l +i(\alpha_{a}^{(I)}-\alpha_{b}^{(J)}) 
			\right)
		\right\} 
	\right. \nn \\
&{}& \left.
	+ \sum_{l=|n_I-n_J|/2}^{(n_I+n_J)/2-1} 
		6(2l+1) \ln \sinh 
		\left\{ \frac{1}{2} 
\left(  \mt \beta (l+\half) +i(\alpha_{a}^{(I)}-\alpha_{b}^{(J)}) 
			\right)
		\right\} 
	\right. \nn \\	
&{}& \left.
	- \sum_{l=|n_I-n_J|/2-\half}^{(n_I+n_J)/2-\frac{3}{2}} 
		4(2l+1) \ln \cosh 
		\left\{ \frac{1}{2} 
			\left( \beta (\frac{l}{3}+\quarter)
			 +i(\alpha_{a}^{(I)}-\alpha_{b}^{(J)}) 
			\right)
		\right\} 
	\right. 
\nn \\
&{}& \left.
	- \sum_{l=|n_I-n_J|/2+\half}^{(n_I+n_J)/2-\half} 
		4(2l+1) \ln \cosh 
		\left\{ \frac{1}{2} 
			\left( \beta (\frac{l}{3}+\frac{1}{12})
			 +i(\alpha_{a}^{(I)}-\alpha_{b}^{(J)}) 
			\right)
		\right\} 
	\right]\, . 
\label{effective-action3} 
\end{eqnarray}
The ``$\ln \, \sinh$'' and ``$\ln \, \cosh$'' terms, which
come from bosonic and fermionic fluctuations, respectively,
yield an attractive potential between the eigenvalues,
while the ``$\ln \, \sin$'' term, which comes from the
Vandermonde determinant, yields a repulsive potential.


\section{Thermodynamic properties of the plane-wave matrix model}
\label{section:thermo-pp}

In order to investigate the
thermodynamic properties of the plane-wave matrix model,
we consider the free energy
defined by 
$F = - T \ln Z - {\rm const.}$, where the constant
\footnote{This constant is the same for all the classical vacua
considered in this paper according to the results of
appendix \ref{section:overall}.}
is subtracted in such a way 
that $\lim_{N\rightarrow \infty}\frac{F}{N^2}$ vanishes at $T=0$. 
This convention is motivated from the requirement
that the free energy $F$ should be of order 1
below the Hagedorn transition \cite{FSS}.
Throughout this section we restrict ourselves to the $s=1$ case,
and use the parameters $n \equiv n_1$ and $k\equiv k_1$.

When we fix $k$ in the large $N$ limit,
the $l = O(N)$ terms 
in the effective action (\ref{effective-action2})
have to contribute in order for the free energy to become of order
$N^2$.
This, however, does not happen unless $\beta = 0$
since for $\beta \neq 0$ the $l \gtrsim 1/\beta$ terms
cancel between bosons and fermions.
Therefore, the free energy 
$\lim_{N\rightarrow \infty}\frac{F}{N^2}$ vanishes at any $T< \infty$,
which implies that 
the Hagedorn temperature is infinite \cite{FSS}.
Note also that there are only a finite number of eigenvalues
in the present case, and hence there are no critical behaviors
associated with the dynamics of the eigenvalues.

On the other hand, when we fix $n$ in the large $N$ limit,
the free energy can be of order $N^2$ at sufficiently high temperature,
since the effective action is generically $O(k^2) \sim O(N^2)$.
When $\beta$ is large,
the attractive potential is insignificant,
and the $\alpha_{a}$ distribute uniformly in the interval
$(-\pi , \pi]$.
In this case, we will see that 
$\lim_{N\rightarrow \infty}\frac{F}{N^2}$ actually vanishes.
As $\beta$ decreases,
the attractive potential becomes more important.
At some point the distribution of $\alpha_{a}$ 
becomes non-uniform and 
$\lim_{N\rightarrow \infty}\frac{F}{N^2}$ becomes negative.

This transition,
which is interpreted as the Hagedorn transition \cite{FSS},
is associated with the spontaneous breakdown of the center symmetry
$A(t) \mapsto A(t) + {\rm const.} {\bf 1}_N$.
The Polyakov line, which is
a useful order parameter for the spontaneous symmetry breaking,
is given, at the leading order of the perturbation theory, by 
\be
P \equiv 
\left\langle 
\frac{1}{N} \left| \tr {\cal P} \exp \left( i
\int_0^\beta  dt \bar{A}(t) \right)  \right| \right\rangle
= \frac{1}{k} \left\langle 
\left| \sum_{a=1}^{k} \exp(i \alpha_a) \right| 
\right\rangle \ ,
\label{polyakov}
\ee
where the expectation value is taken with respect to the
effective action (\ref{effective-action2})
for the gauge field moduli.
When the center symmetry is unbroken and 
the distribution of $\alpha_a$ is uniform, we obtain
$ P = 0$\, in the large $N$ limit.
When the center symmetry is spontaneously broken,
and the distribution of $\alpha_a$ is non-uniform, we obtain
$ P \neq 0$\,.
This is analogous 
to the deconfinement phase transition 
in large $N$ gauge theories 
\cite{AW,Sundborg,Aharony1,Aharony2,SV,Witten:thermal}.

In what follows we consider the $n = 1,2,3$ cases, 
for which $k = \frac{N}{n}$ goes to infinity with $N$.
The integration over the gauge field moduli,
which play a crucial role in the Hagedorn transition,
can be done analytically near the transition point.
For arbitrary temperature we perform
Monte Carlo simulation to confirm the analytical results and
to obtain explicit results in the high temperature regime, 
which is not accessible analytically. 

\subsection{Analytical results near the transition point}
\label{section:near-transition}

In this section, we investigate the behavior
near the Hagedorn transition analytically
using the eigenvalue density
\be
\rho(\theta)=\frac{1}{N}\sum_{a=1}^{k}
\delta(\alpha_a-\theta)\ ,
\label{eigenvalue-density}
\ee
where $\theta \in ( - \pi ,  \pi ] $.

First let us review the analysis for the trivial vacuum
given in refs.\ \cite{FSS,Semenoff}.
The effective action (\ref{effective-action1}) can be written
in terms of the eigenvalue density $\rho(\theta)$ as
\be
\label{effective-action4}
S_{\rm eff} =
N^2 \int \int d\theta d\theta' \rho(\theta) \rho(\theta') 
\left\{ - \sum_{m=1}^{\infty} \frac{1}{m} 
f_m(\beta)  e^{im(\theta-\theta')} -
\ln \left| \sin 
\left(  \frac{1}{2} (\theta - \theta') \right)
\right| \right\} \, , 
\ee
where we have expanded the 
``$\ln\sinh$'', ``$\ln\cosh$'' in (\ref{effective-action2})
into Fourier series,
and the function $f_m(\beta)$ is defined by 
\be
f_m(\beta) \equiv 3e^{-\frac{1}{3}\beta m} 
+ 6e^{-\frac{1}{6}\beta m} - 8(-1)^m e^{-\frac{1}{4}\beta m} \, . 
\ee
By making a Fourier expansion for the eigenvalue density as
\be
\rho(\theta)=\frac{1}{2\pi} \sum_{m=-\infty}^{\infty} 
\tilde{\rho}_m e^{im\theta} \ , 
\label{rho-m}
\ee
the effective action can be rewritten as
\footnote{In arriving at (\ref{rho-effective}),
we have also expanded the ``$\ln \sin$'' term using the formula
\beq
\ln \sin \frac{\theta}{2} =
- \sum_{m=1}^\infty \frac{1}{m} \cos m \theta + {\rm const}.
\eeq
}
\be
S_{\rm eff}= N^2 \sum_{m=1}^{\infty}
\frac{1}{m}
|\tilde{\rho}_m|^2\Bigl\{ 1-f_m(\beta) \Bigr\} \ .
\label{rho-effective}
\ee
If one of the coefficients of $|\tilde{\rho}_m|^2$
becomes negative, the uniform distribution becomes unstable.
This indeed occurs for $m=1$ as we decrease $\beta$ from $\infty$. 
The Hagedorn temperature $T_H \equiv 1/\beta_H$, 
which can be determined by solving  
\be
1 - f_1(\beta_H)=0 \ , 
\label{hagedorn-temperature}
\ee
is $T_H \simeq 0.0758533$.

Since the $\beta_H = 1/T_H $ turned out to be quite large,
we may safely omit the $m \ge 2$ terms in the expansion
(\ref{effective-action4}).
This simplification enables the analytic calculation
of the thermodynamic quantities by the saddle-point method, 
which is exact in the large $N$ limit.
The saddle-point equation for the eigenvalue density reads
\be
0 = \frac{1}{N^2} \frac{d}{d \theta}
 \frac{\delta S_{\rm eff} }{\delta \rho(\theta)}
 \simeq \int d\theta' \rho(\theta') \left\{
f_1 (\beta) \sin(\theta-\theta') - 
\frac{1}{2}\cot\left(\half (\theta-\theta')\right)\right\} \ .
\label{sad-point-eq}
\ee
This equation can be solved by using the Ansatz of the
Gross-Witten form \cite{GW}
\be
\rho(\theta) =   \left\{
\begin{array}{cl}
\frac{2}{\pi\omega} \cos\frac{\theta}{2} 
	\sqrt{\frac{\omega}{2}-\sin^2\frac{\theta}{2}}
& \left( | \theta | \le \theta_{\rm cl} \right) \, , \\
0 & ( |\theta| > \theta_{\rm cl} ) \, , \\
\end{array}
\right. 
\label{eigenvalue-distribution_exact}
\ee
where $\theta_{\rm cl} = 2\sin^{-1} \sqrt{\frac{\omega}{2}}$ and
\be
\omega = 2\left(1-\sqrt{1-\frac{1}{f_1(\beta)}}\right) \ .
\label{lambda-def}
\ee
Given the eigenvalue density, the free energy
\footnote{We have corrected an error in the analytic expression 
for the free energy given in ref.\ \cite{Semenoff}.} 
and the Polyakov line can be calculated as (See
Appendix \ref{section:eval-free} for the derivation.)
\beqa
\frac{F}{N^2} 
&=&  - \frac{T S_{\rm eff}}{N^2} 
\simeq
- T \left\{\frac{f_1(\beta)}{2} \left(2-\frac{\omega}{2}
\right)
+\half \ln \left(\frac{\omega}{2} 
\right) - \half \right\} \ ,
\label{free-energy_exact} \\
P &=& \int d\theta \rho(\theta) e^{i\theta}
\simeq
1 - \frac{\omega}{4}
\, .
\label{polyakov_exact}
\eeqa

The above analysis can be extended to the general $n$ case 
in a straightforward manner.
The only modifications 
in the effective action (\ref{effective-action4})
are to replace the overall factor $N^2$ 
by $k^2$, and to replace the functions $f_m(\beta)$ by 
\bear
f_m(\beta) &=& \sum_{l=0}^{n-2} (2l+1) e^{-\mt \beta m (l+1)} 
	+ \sum_{l=1}^{n} (2l+1) e^{-\mt \beta m l} 
	+ \sum_{l=0}^{n-1} 6(2l+1) e^{-\mt \beta m (l+\half)} \nn \\
& {} & \hspace{0cm}
	- \sum_{l=\half}^{n-\frac{3}{2}} (-1)^m 
		4(2l+1) e^{- \beta m (\frac{l}{3}+\frac{1}{4})}
 	- \sum_{l=\half}^{n-\half} (-1)^m
	4(2l+1) e^{- \beta m (\frac{l}{3}+\frac{1}{12})} \ . 
\label{hagedorn-temperature2} 
\eear
By solving eq.\ (\ref{hagedorn-temperature}) in the present case,
we obtain the Hagedorn temperature 
\be
T_H \simeq \left\{
\begin{array}{ll}
0.0738901  & \hspace{0.5cm} (n = 2) \\
0.0738526 & \hspace{0.5cm} (n = 3) \\
0.0738520  & \hspace{0.5cm} (n \geq 4) \ . 
\end{array}
\right.
\label{hagedorn-temperature3}
\ee
Since these values are small,
we may neglect the $m \ge 2$ terms in (\ref{effective-action4})
in the general case as well.
The expression for the free energy (\ref{free-energy_exact})
gets multiplied by the factor $\frac{1}{n^2}$
due to the modified prefactor in $S_{\rm eff}$ mentioned above,
while the expression for the Polyakov line (\ref{polyakov_exact})
remains the same.
Note that 
one has to use $f_1 (\beta)$
given by (\ref{hagedorn-temperature2})
for the definition of $\omega$
in eq.\ (\ref{lambda-def}).
In section \ref{section:monte}
we will confirm the validity of these analytical results
by Monte Carlo simulation.

\subsection{High temperature limit} 
\label{section:highTlim}

At high temperature (i.e., small $\beta $),
the ``$\ln\sinh$'' terms in the effective action (\ref{effective-action2})
make the eigenvalues attracted
to each other against the repulsive force coming from
the ``$\ln\sin$'' terms,
and therefore $(\alpha_a - \alpha_b)$ is typically
of order $\beta$.
The free energy can therefore be estimated by simply replacing
the ``$\ln\sinh$'' and ``$\ln\sin$'' terms by $\ln \beta$,
and omitting the ``$\ln\cosh$'' terms in the effective action
(\ref{effective-action2}).
This gives
\be
\frac{F}{N^2} = - \left( 8  + \frac{1}{nN} \right) 
T \ln T + O(T) \ .
\label{free-energy_large-T}
\ee
Thus, the leading asymptotic behavior of 
$ \lim_{N \rightarrow \infty} \frac{F}{N^2}$
at high temperature is universal.
In section \ref{section:monte} we will confirm this asymptotic
behavior by Monte Carlo simulation.

\subsection{Monte Carlo integration over the gauge field moduli}
\label{section:monte}

In this section we 
perform Monte Carlo simulation
to integrate over the gauge field moduli,
and obtain explicit results 
for the free energy and the Polyakov line
at arbitrary temperature.

\FIGURE[htb]{\epsfig{file=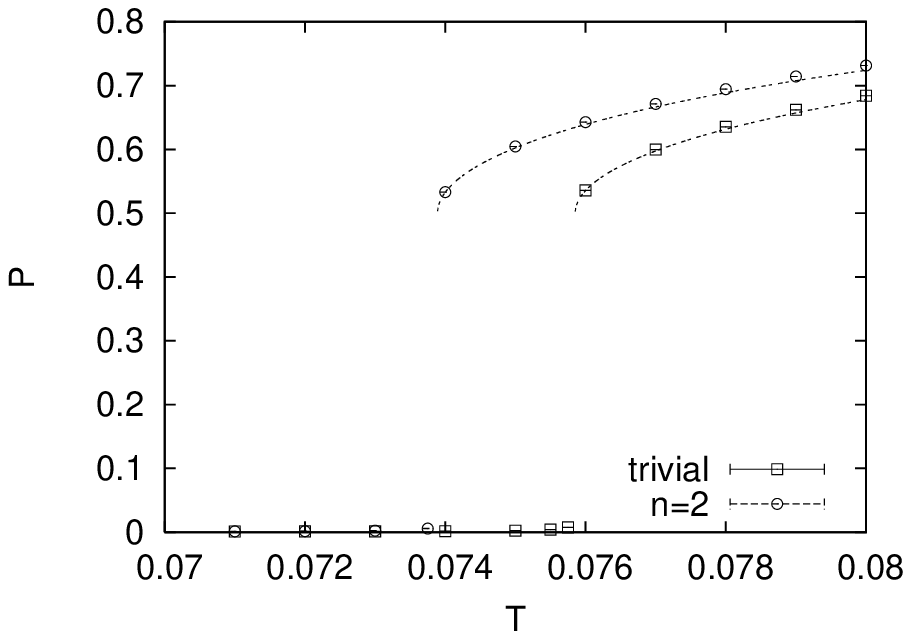, width=.49\textwidth}
\caption{The Polyakov line
$P$ is plotted against the temperature $T$ 
near the critical point 
for $N = 2000$ in the trivial case, 
and for $k=2000$ in the $n=2$ case.
The dotted lines represent the 
analytic results (\ref{polyakov_exact}).
}
\label{polyakov_data}}

We use the Metropolis algorithm for the simulation. 
At each step we generate 
a trial configuration by replacing one of the eigenvalues
by a random number within the interval 
$(- \pi, \pi]$,
and accept it with the probability
$ \, \max (1, e^{-\Delta S_{\rm eff}}) \, $, where
$\Delta S_{\rm eff}$ is the difference of the effective action 
(\ref{effective-action2}) for the trial configuration 
from that for the previous configuration.

In Fig.\ \ref{polyakov_data}
we plot the expectation value of the Polyakov line
near the transition point.
Our Monte Carlo results are in good agreement 
with the analytic results (\ref{polyakov_exact})
including the position of the transition point.
We have also made a similar plot for the $n=3$ case,
but it turned out to be almost indistinguishable from the
results for the $n=2$ case.
In Figs.\,\ref{density_trivial} and \ref{density_2sphere}
we plot the distribution of the eigenvalues.
While our data agree with the analytic results 
(\ref{eigenvalue-distribution_exact})
near the transition point, we also start to see a small deviation
as the temperature increases.


\DOUBLEFIGURE[htb]{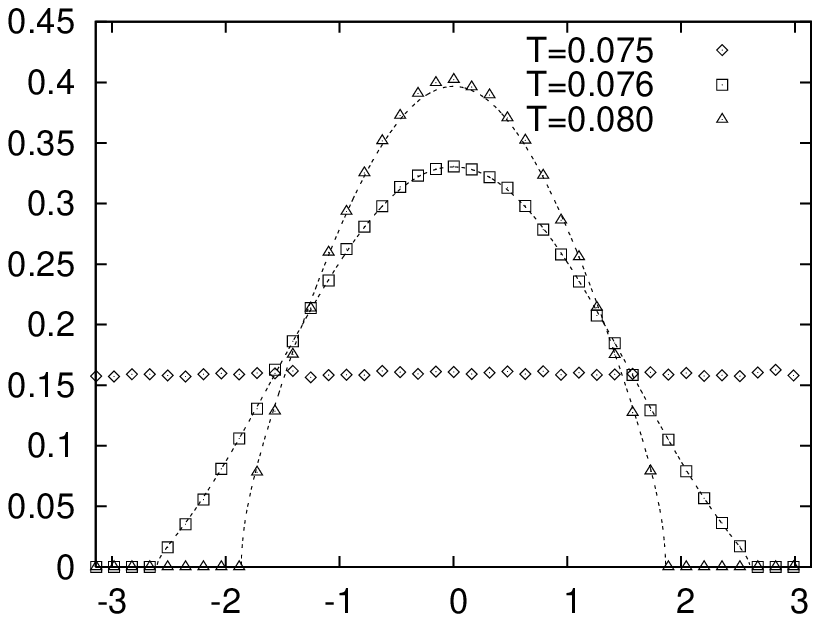, width=.49\textwidth}
{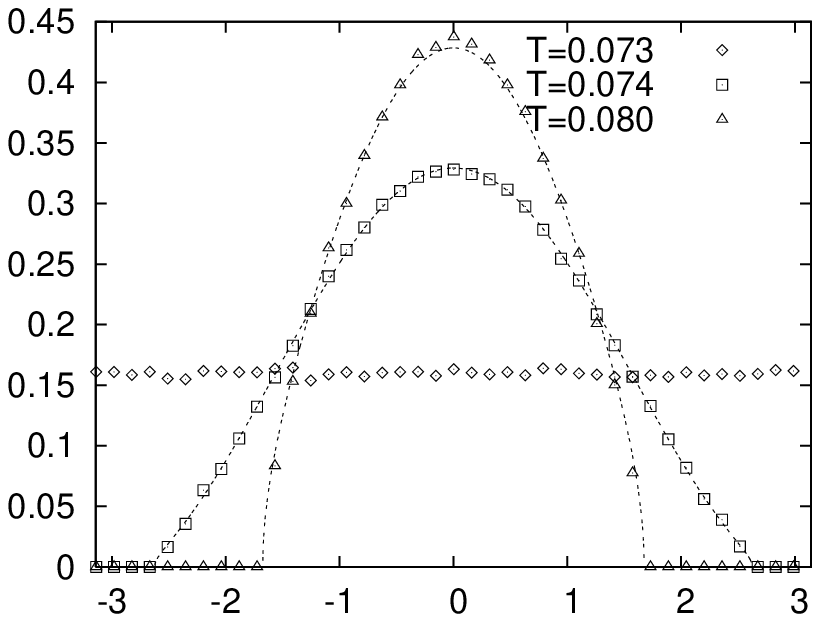, width=.49\textwidth}
{The eigenvalue distribution $\rho(\theta)$ 
in the trivial vacuum case
is plotted 
for $T=0.075 $, $0.076 $, $0.080 $ and $N=2000$\,. 
The dashed lines represent the 
analytic result (\ref{eigenvalue-distribution_exact}) 
obtained with the Ansatz of the Gross-Witten form.
\label{density_trivial}}
{
The eigenvalue distribution $\rho(\theta)$ in the $n=2$ case
is plotted for 
$T=0.073 $, $0.074 $, $0.080 $ and $k=2000$\,. 
The dashed lines represent the 
analytic result (\ref{eigenvalue-distribution_exact}) 
obtained with the Ansatz of the Gross-Witten form.
\label{density_2sphere}}

The free energy can be calculated by
\be
F = - T \langle S_{\textrm{eff}} \rangle
\label{free-energy2}
\ee
in the large $N$ limit.
In order to obtain the free energy accurately near the
critical point, we have to make a large $N$ extrapolation.
The dominant finite $N$ effects come from the 
``$\ln \sin$'' term in the effective action
(\ref{effective-action2}) due to its logarithmic singularity
when the eigenvalues come close to each other.
Since the distance between the nearest eigenvalues is 
of order $\frac{1}{N}$, the finite $N$ effects for evaluating
the $O(1)$ quantity
\be 
\frac{1}{N^2} \sum_{a
\ne b}^{N} \ln \left| \sin \frac{\alpha_a - \alpha_b}{2} 
\right| \, , \nn
\ee
for instance, is $O\left( \frac{\ln N}{N} \right) \,.$
In Fig.\,\ref{free_energy_finite_N}
we therefore plot the free energy obtained
by (\ref{free-energy2}) at finite $N$
against $\frac{\ln N}{N}$
in the trivial vacuum case for various temperature.
Indeed our data can be nicely fitted
to straight lines, from which we can make a reliable
large-$N$ extrapolation.
The large-$N$ limits obtained in this way are plotted 
in Fig. \ref{free_energy_large_N}
against the temperature $T$ near the critical point.
Our results agree nicely to 
the analytic results (\ref{free-energy_exact}). 
At sufficiently high temperature, we observe that
the trivial vacuum gives
the smallest free energy.
As we decrease the temperature, however, it is taken
over by the $n=2$ vacuum.

\DOUBLEFIGURE[htb]
{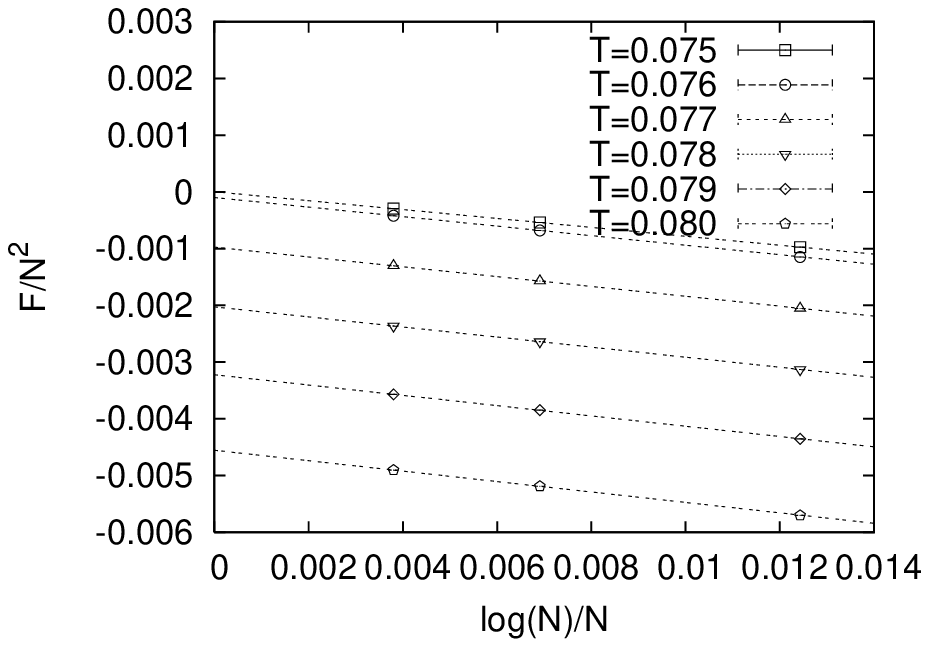, width=.49\textwidth}
{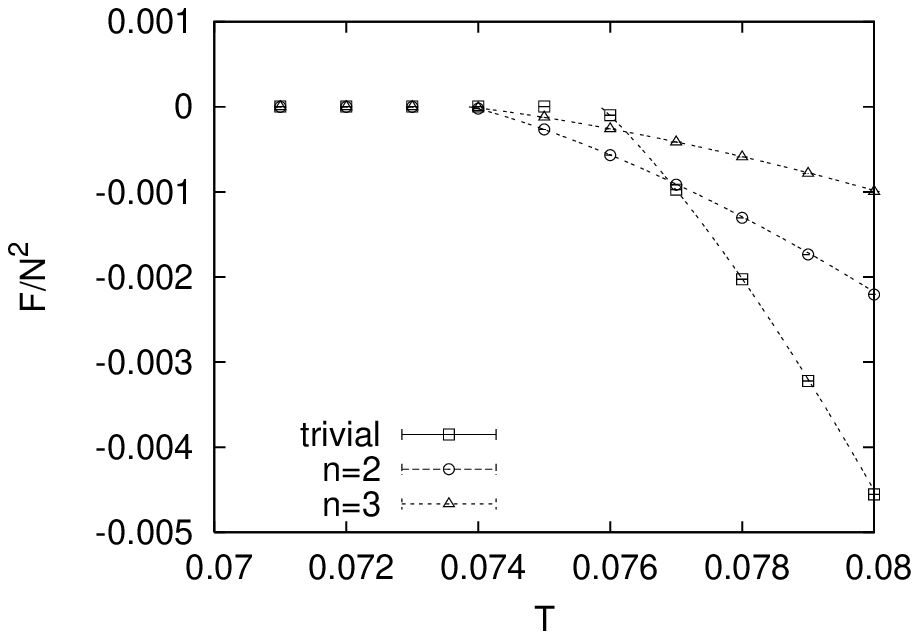, width=.49\textwidth}
{The free energy $F$ 
is plotted against $\frac{\ln N}{N}$ 
for $N=500, \, 1000$ and $2000$ in the trivial vacuum case
for various temperature.
Our data are nicely fitted to straight lines.
\label{free_energy_finite_N}}
{The free energy $F$ obtained by the large $N$ extrapolation
is plotted against the temperature $T$ near the critical point
for the trivial vacuum as well as for the $n=2,3$ cases.
The curves represent the analytic results (\ref{free-energy_exact}).
\label{free_energy_large_N}}





In Figs.\ \ref{high_trivial} and \ref{high_sphere}
we plot the free energy $\frac{F}{N^2}$ at higher temperature.
The deviation from
the analytic result for
the free energy (\ref{free-energy_exact})
becomes pronounced for $T \gtrsim 1$ as expected,
and our data can be nicely fitted to
\be
\frac{F}{N^2} \simeq -8T \log(T) - c_1 T - c_2  \ ,
\label{highT-fitting}
\ee
where the leading asymptotic behavior is determined analytically
in section \ref{section:highTlim}.
Although the leading term is the same at $N=\infty$,
the coefficient $c_1$ of the linear term turns out
to be larger for the trivial vacuum case.
Hence we conclude that the trivial vacuum gives the smallest
free energy for $T \gtrsim 0.077$.

\DOUBLEFIGURE[htb]
{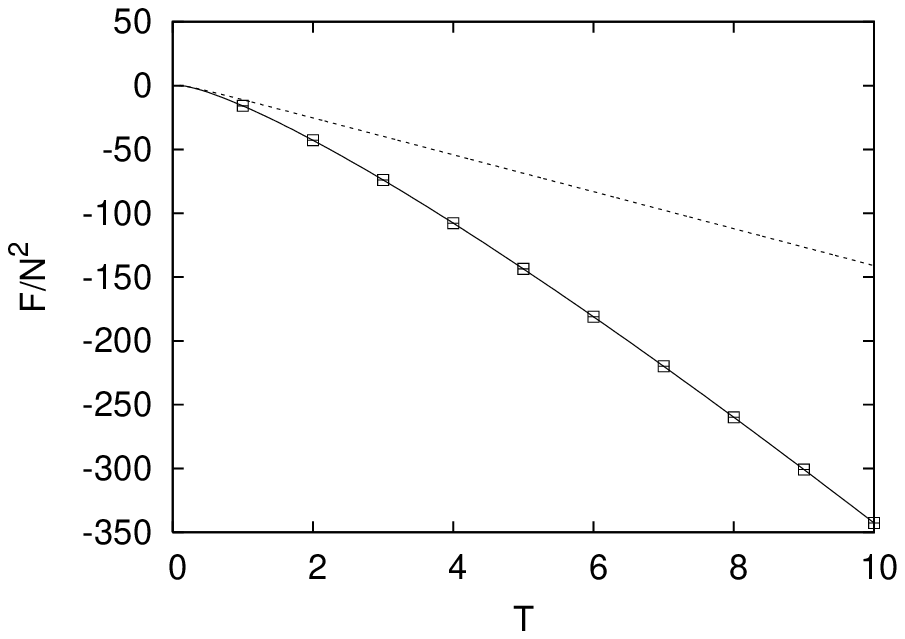, width=.49\textwidth}
{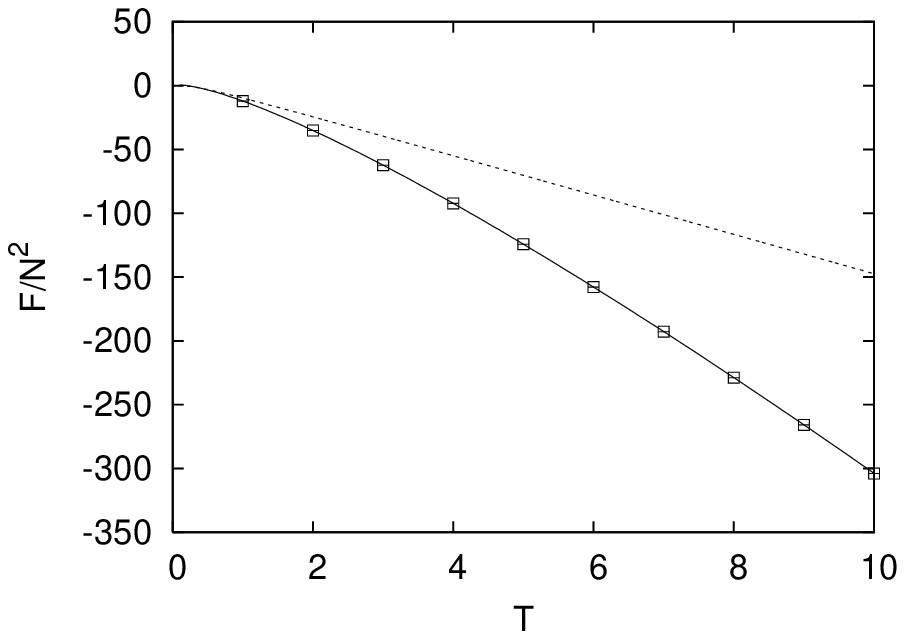, width=.49\textwidth}
{The free energy $\frac{F}{N^2}$ 
is plotted against $T$ 
for $N=500$ in the trivial vacuum case.
The solid line represents the result for fitting
the data for $1\le T \le 10$ to
(\ref{highT-fitting}),
where $c_1 = 15.852(1)$ and $c_2 = 0.038(3)$.
The dotted line represents the analytic result for
the free energy (\ref{free-energy_exact}).  
\label{high_trivial}}
{
The free energy $\frac{F}{N^2}$ 
is plotted against $T$ 
for $n=2$ and $k=500$. 
The solid line represents the result for fitting
the data for $1\le T \le 10$ to
(\ref{highT-fitting}),
where 
$ c_1 = 11.948(5)$ and $c_2 = 0.16(1)$.
The dotted line represents the analytic result for
the free energy (\ref{free-energy_exact}).  
\label{high_sphere}}

\section{Summary and discussions} 
\label{section:summary}

We have studied the thermodynamic properties
of the plane-wave matrix model in general classical vacua.
The Hagedorn transition occurs due to the
dynamics of the gauge field moduli.
Extending the previous works on the trivial vacuum,
we obtained analytical results for the thermodynamic quantities
near the transition point,
which revealed the existence of the ``fuzzy-sphere'' phase.
We also performed Monte Carlo simulation to integrate over the
gauge field moduli. This method can be used for arbitrary temperature.
Usually free energy is difficult to calculate by Monte Carlo simulation,
but in the present case we can obtain it very accurately from the
expectation value of the effective action for the gauge field moduli,
thanks to large $N$.
We observe that the trivial vacuum gives the smallest free energy
in the high temperature regime.


The plane-wave matrix model is closely related to $\mathcal{N}=4$ SYM
theory on $R\times S^3$ \cite{KKP}, which implies that the matrix
model at finite temperature is related to the
SYM theory
on $S^1\times S^3$\,. We therefore expect that our results should have
implications on the phase structure of the $SU(\infty)$ SYM in four
dimensions. In particular, the fuzzy sphere phases we found in this
paper may correspond to the plasma-ball phases in the
$SU(\infty)$
SYM on $S^1\times S^3$, which
are interpreted as localized black holes on the gravity side via the
AdS/CFT duality \cite{Aharony:2005cx}. 
Another interesting future direction
is to clarify the relation between our result and the
entropy bound discussed recently \cite{Bousso}. 




We hope that our work provides a clue to
the phase structure of the AdS black hole 
and to a deeper understanding of
the AdS/CFT correspondence at finite temperature.

\vspace*{0.5cm}
\section*{Acknowledgments} 

We would like to thank T.~Azuma, H.~Fuji,
Y.~Hikida, S.~Mizoguchi, Y.~Susaki, 
S.~Takeuchi and A.~Yamaguchi for helpful discussion,
and G.~Semenoff for explaining his works on this subject.
K.Y.\ is grateful to H.~Shin for a series of fruitful collaborations, 
which are closely related to the topic discussed in this paper. 
The work of K.Y.\ is supported in part by JSPS
Research Fellowships for Young Scientists.

\vspace*{1cm}
\appendix

\section*{Appendix}

\section{Gauge fixing}
\label{section:gauge-fixing}

In order to arrive at our main result
eq.\ (\ref{partition-fn-oneloop})
in section \ref{section:one-loop},
we have first integrated over the fluctuations
fixing the gauge field moduli, and then the
integration over the gauge field moduli has been
written in terms of the moduli parameters $\alpha^{(I)}_{a}$.
Since the classical vacua (\ref{fuzzy-sphere1}) 
breaks the $U(N)$ gauge symmetry 
down to $\prod_{I=1}^{s}U(k_I)$, we have to make gauge fixing
at each step of the above procedure.
First, when we integrate over the fluctuations,
we have to fix the gauge along the gauge orbit
in the direction of the coset space
$\frac{U(N)}{\prod_{I=1}^{s}U(k_I)}$.
This is necessary since otherwise there will be zero modes
corresponding to the broken symmetry, and we cannot integrate
over the fluctuations.
Next, when we integrate over the gauge field moduli,
we have to fix the gauge corresponding to the remaining symmetry
$\prod_{I=1}^{s}U(k_I)$ in order to reduce the integration
to that over the moduli parameters $\alpha^{(I)}_{a}$.
This is necessary for studying the large $N$ limit analytically
by the eigenvalue density method and for simplifying
Monte Carlo calculations drastically.
In what follows we explain the two steps of gauge fixing procedures
separately.

\subsection{Integrating the fluctuations}
\label{section:gf-fluctuation}

Let us explain the first gauge fixing,
which is necessary for integrating the fluctuations.
We use the background field gauge 
\begin{eqnarray}
\overline{D}_\mu A_\mu (t) = r(t) \ ,
\label{gauge-fixing1} 
\end{eqnarray} 
where $r(t)$ is an arbitrary function,
and $\overline{D}_\mu$ and $A_\mu (t)$ for $\mu = t,1,2,3$
are defined by
\beqa
&& \overline{D}_t = \dt  - i[\bA(t), \  \cdot \ ] \, , \quad
A_t (t) = A(t) \, , \\
&& \overline{D}_i =  - i[B_i, \  \cdot \ ]  \, , \quad\quad
A_i (t) = X_i (t) \quad\quad
\mbox{for $i=1,2,3$} \, .
\eeqa
The corresponding Faddeev-Popov (FP) determinant is defined by
\begin{eqnarray}
1=\int dg(t) \delta
\left(\overline{D}_{\mu} A_{\mu}^{g}(t) - r(t)\right)
\Delta_{\rm FP}^{(1)} \ , 
\label{identity1}
\end{eqnarray} 
where $g(t)$ is an element of the coset group
$\frac{U(N)}{\prod_{I=1}^{s}U(k_I)} $
and $A_\mu ^{g}(t)$ represents the gauge transformed field.
The FP determinant can be represented as
\begin{eqnarray}
\Delta_{\textrm{FP}}^{(1)}
= \int [dc][d\bar{c}] \exp
\left(-\int\!dt\, 
\tr
\left(\bDt \bar{c} \cdot \Dt c -
[B_i,\bar{c}][X_i,c]\right) \right) \ ,
\label{FP1}
\end{eqnarray}
where $c$, $\bar{c}$ are the ghost fields 
satisfying the periodic boundary condition. 
Inserting the identity (\ref{identity1})
in the partition function,
and integrating over $r(t)$ with the gaussian weight
$\exp \left(-\half \int\!dt\, r^2(t) \right)$,
we obtain the gauge fixing term and the ghost term
\begin{eqnarray}
S_{\rm g.f.} =  \int\! dt\,  \tr \left\{ 
\half \left( \overline{D}_{\mu} A_{\mu} \right)^2 
\right\} \, ,
\quad S_{\rm ghost} =
\int\! dt\,  \tr \left\{ 
\bDt \bar{c} \cdot \Dt c 
- [B_i,\bar{c}][X_i,c] \right\}\ ,
\label{gf-ghost-terms}
\end{eqnarray}
which should be added to the original action (\ref{BMN-action}).

\subsection{Integrating the gauge field moduli}
\label{section:gf-moduli}

Let us explain the second gauge fixing, which is necessary
for integrating the gauge field moduli. We use the
static diagonal gauge, which is analogous to the one adopted
for the trivial vacuum \cite{FSS}. 

First we impose the static gauge condition given by
\begin{eqnarray}
\dt \bar{A}^{(I)}(t) = 0 \ . 
\label{gauge-fixing2} 
\end{eqnarray} 
The corresponding FP determinant can be defined by
\be
1=\int \prod_{I=1}^{s} d \tilde{g}^{(I)}(t) 
\delta \left( \frac{d \bA ^{(I)}(t)}{dt} \right) \Delta_{\rm FP}^{(2)}\ , 
\label{identity2}
\ee
where $\tilde{g}^{(I)}(t)$ is an element 
of the unbroken gauge group $\prod_{I=1}^{s}U(k_I)$.
The FP determinant $\Delta_{\textrm{FP}}^{(2)}$ is given by
\begin{eqnarray}
\Delta_{\textrm{FP}}^{(2)} 
= \prod_{I=1}^{s} \det{}' \left( \dt\bDt^{(I)} \right)\ , 
\label{FP2}
\end{eqnarray}
where we have defined
$\bDt^{(I)} \equiv \dt - i[\bA^{(I)}, \ \cdot \ ]$, and 
the symbol $\det '$ implies that we have omitted the zero modes.
The static gauge (\ref{gauge-fixing2}) does not fix the gauge completely,
and we still have the global $\prod_{I=1}^{s}U(k_I)$ symmetry
as the residual gauge symmetry, which we fix by further imposing the
constant matrices $\bar{A}^{(I)}_{ab}$ to be 
diagonal (\ref{gauge-fixing3}).


As is well known in matrix models, this gauge fixing yields
the Vandermonde determinant, which is derived as follows.
The path integral measure for the
gauge field moduli $[d \bA (t)]$
around the static diagonal configuration 
(\ref{gauge-fixing3}) is rewritten as  
\begin{eqnarray}
d \bar{A}^{(I)}_{ab}=
\prod_{I=1}^{s} 
\left( \frac{1}{k_I!} 
\cdot \prod_{a=1}^{k_I} \frac{d\alpha^{(I)}_{a}}{2\pi} 
\cdot \prod_{a \ne b}^{k_I}
\left| \alpha^{(I)}_{a}-\alpha^{(I)}_{b} \right| 
\cdot dg^{(I)} 
\cdot \prod_{m \ne 0}
\prod_{a,b}^{k_I} d\bA^{(I)}_{m, ab} \right)\ , 
\label{measure}  
\end{eqnarray}
where $g^{(I)}$ is an element of 
$\prod_{I=1}^{s}U(k_I)$,
and $d\bA^{(I)}_m$ ($m \ne 0$) is the integration measure for
the non-zero Fourier modes.
%



Let us evaluate the determinant
(\ref{FP2}) explicitly.
In terms of Fourier modes
\footnote{The fields $M(t)$ and $\Theta(t)$ obeying
the periodic and anti-periodic boundary conditions, 
are expanded as
\begin{eqnarray}
M(t) = \frac{1}{\sqrt{\beta}} \sum_{m=-\infty}^{\infty} M_m
{\rm e}^{\frac{2\pi i m}{\beta} t}\ , \qquad   
\Theta(t) = \frac{1}{\sqrt{\beta}} \sum_{m=-\infty}^{\infty}
	\Theta_m {\rm e}^{\frac{\pi i (2m-1)}{\beta} t}\, , 
\label{fourier}
\end{eqnarray}
respectively.},
the FP determinant (\ref{FP2}) reads
\be
\Delta_{\textrm{FP}}^{(2)} 
= \prod_{I=1}^{s} 
	\left[ 
	\prod_{a , b=1}^{k_I} \prod_{m \ne 0} 
	(2\pi imT) \Bigl\{ 2\pi i m T 
	+ i (\alpha_{a}^{(I)}-\alpha_{b}^{(I)} ) T \Bigr\}
	\right] \, .
\ee
Therefore the identity (\ref{identity2}) is rewritten as
\begin{eqnarray}
1 &=& \prod_{I=1}^{s} 
	\left[ d{g'}^{(I)}(t)
		\prod_{a,b=1}^{k_I} \prod_{m \ne 0} 
		\delta(2\pi i m T\bA^{(I)}_m) (2\pi i m T)
\Bigl\{ 2\pi i m T + i (\alpha_{a}^{(I)}-\alpha_{b}^{(I)} ) T
\Bigr\}	\right]  \nn \\
&=& \prod_{m \ne 0} \left( 2\pi i m T \right)^{\sum_{I=1}^{s}k_I^2}
\cdot \prod_{I=1}^{s}
		\left[ d{g'}^{(I)}(t) \prod_{a \ne b}^{k_I} 
			\prod_{m=1}^{\infty} 
\left\{ 1 - \left( \frac{
\alpha_{a}^{(I)}-\alpha_{b}^{(I)} 
}{2\pi m}\right)^2 
			\right\}
		\right] \ . 
\label{deldel}
\end{eqnarray}
In the second line, we have omitted 
the delta function for the non-zero Fourier modes,
which are integrated out by using the measure 
in (\ref{measure}). 
{}Inserting the identity (\textrm{\ref{deldel}}) in the partition
function, the measure (\textrm{\ref{measure}}) 
for the gauge field moduli $[d \bA(t)]$ can be written as
\beqa
[d \bA(t)] &=&
\left( 2T \right)^{\sum_{I=1}^s k_I^2} 
\cdot \prod _{m \ne0 } \left(2\pi i mT \right)^{\sum_{I=1}^s k_I^2} 
\\
&~& \times \prod_{I=1}^{s} 
\left\{ dg^{(I)}(t)
	\cdot \frac{1}{k_I!} 
\prod_{a=1}^{k_I} \frac{d\alpha_{a}^{(I)}}{4\pi T}
	\cdot \prod_{a \ne b}^{k_I} 
\left|	\sin
\frac{\alpha_{a}^{(I)}-\alpha_{b}^{(I)} }{2}
\right| 
\right\}\ , \label{mes-sine}
\eeqa 
where $dg^{(I)}(t) \equiv d\tilde{g}^{(I)}(t) \cdot dg^{(I)}$
represents the measure for the gauge function 
of the $\prod_{I=1}^{s}U(k_I)$ symmetry. 
The ``$\sin$'' term in (\ref{mes-sine}) appeared
from the formula used to sum over the Fourier modes
\begin{equation}
\prod_{n=1}^{\infty} \left( 1 - \frac{x^2}{n^2} \right) 
= \frac{\sin \pi x}{\pi x} \nn \, .
\end{equation}

The result for the trivial vacuum can be obtained formally by
setting $s =1$\,, $k_1 = N$ and $n_1 =1$\,, which yields
\begin{eqnarray}
\left( 2T \right)^{N^2} 
\cdot \prod _{m \ne 0} \left(2\pi i mT \right)^{N^2}
\cdot dg(t) \cdot \frac{1}{N!} \,
	\prod_{a=1}^{N} \frac{d\alpha_{a}}{4\pi T}
	\cdot \prod_{a \ne b}^{N} 
	\left| \sin 
\frac{\alpha_{a}-\alpha_{b}}{2}
\right|  \, ,
\end{eqnarray}
where $g(t)$ represents the gauge function
of the $U(N)$ symmetry.
This agrees with the measure obtained in ref.\ \cite{FSS}.

\section{Evaluation of the determinant}
\label{section:eval-det}

Integrating over the fluctuation $Y$, $\Psi$, $\tA$ and ghost fields
using the quadratic action (\ref{quadS-rewritten}),
we obtain the determinants
\bear
Y, \tA &:& \quad  \det{}_{\rm B}^{-1/2} (\bDt^2 + \lambda^2)\ ,  \nn \\
\Psi &:& \quad \det{}_{\rm F} (\bDt + \lambda)\ ,  \nn \\ 
c, \bar{c} &:& \quad \det{}_{\rm B} (\bDt^2 + \lambda^2)\ ,
\eear
where $\lambda$ represents an eigenvalue of the mass operators
(\ref{mass-operator}), and the lower suffices of $\det$ specify 
the boundary condition (periodic for ``B'' and anti-periodic for ``F'')
for the fields on which the operators act.

Using the formulae
\be
\prod_{n=1}^{\infty} \left( 1 + \frac{x^2}{n^2} \right) 
= \frac{\sinh \pi x}{\pi x}\ , \qquad
\prod_{n=1}^{\infty} \left( 1 + \frac{x^2}{(2n-1)^2} \right) 
= \cosh \left(\frac{\pi x}{2}\right) \nn \, , 
\ee
we obtain the determinants as
\begin{eqnarray}
\det{}_{\rm B}^{1/2} \left( -\bDt^2 + \lambda^2 \right) 
&=& \prod_{m} \prod_{a,b}
		\left\{ 2\pi i mT + i (\alpha_{a}-\alpha_{b})T
		+ \lambda \right\} \nn \\
&=& (2T)^{N^2} \cdot \prod_{m \ne 0} \left( 2\pi i m T \right)^{N^2} 
		\cdot \prod_{a,b} \sinh
	\left\{\frac{1}{2} 
	\left( \beta \lambda  + i (\alpha_{a}-\alpha_{b})
 \right) \right\}\, , \\
\det{}_{\rm F} \left( \bDt + \lambda \right) 
&=& \prod_{m} \prod_{a,b}
		\left\{
		\pi i (2m-1) T + i (\alpha_{a}-\alpha_{b} ) T
		+\lambda \right\} \nn \\
&=& \prod_{m} \left( \pi i (2m-1) T \right)^{N^2} 
\cdot \prod_{a,b} \cosh \left\{\frac{1}{2} 
\left( \beta \lambda  + i (\alpha_{a}-\alpha_{b} )
 \right)
 \right\}\ . 
\end{eqnarray}
The coefficients in front of ``$\sinh$'' and ``$\cosh$''\,
have to be taken into account in comparing the free energy
for different vacua.

\section{The overall factor of the partition function}
\label{section:overall}

In this section we obtain the overall factor
$\mathcal{C}$ in (\ref{partition-fn-oneloop}).
There are three types of contributions to $\mathcal{C}$, 
which affect the calculation of free energy at the leading order in $N$;
namely the Gaussian integration, the determinants,
and the gauge volume obtained from gauge fixing.
The factor obtained from each contribution is 
given in table
\ref{factor2}. 
The symbol $\textrm{Vol}(U(N)){}_{\textrm{local}}$ 
represents the gauge volume for the local $U(N)$ symmetry,
and we have introduced the functions
\begin{eqnarray}
f(T) \equiv 2T \cdot \prod_{m \ne 0} \left( 2\pi i m T \right)\ , \quad 
g(T) \equiv \prod_{m} \Bigl\{ \pi i (2m-1) T \Bigr\} \ . \nn 
\end{eqnarray}
\TABLE[ht]{\footnotesize 
\begin{tabular}{|c|c|c|c|}	\hline
{} & Gaussian & determinant & gauge volume \\ \hline\hline
$\int dg(t)$ & {} & {} & 
$\left(\frac{\textrm{Vol}(U(N))}
{\prod_{I=1}^{s} \textrm{Vol}(U(k_I))}\right)_{\textrm{local}}$ \\ \hline
$\int \prod_{I}^{s} dg{}^{(I)}(t)$ & {} & {} &
$\prod_{I=1}^{k_I}\textrm{Vol}(U(k_I))_{\textrm{local}}$ \\ \hline 
$r(t)$-int. & 
$\left(\prod_{m}{2\pi}\right)^{- \half (N^2-\sum_{I=1}^{s}k_I^2)}$ &
{} & {} \\ \hline 
$Y(t)$-int. & 
$\left( \prod_m 2\pi \right)^{\frac{9}{2}N^2}$ &
$f(T)^{-9N^2}$ & 
{} \\ \hline
$\tA(t)$-int. & 
$\left( \prod_m 2\pi \right)^{\frac{1}{2}(N^2-\sum_{I=1}^{s}k_I^2)}$ &
$f(T)^{-(N^2-\sum_{I=1}^{s}k_I^2)}$ &
{} \\ \hline 
$\Psi(t)$-int. & {} & 
$g(T)^{8N^2}$ &
{} \\ \hline
$c(t)$, $\bar{c}(t)$-int. & {} &
$f(T)^{2(N^2-\sum_{I=1}^{s}k_I^2)}$ &
{} \\ \hline 
$[d \tA]$ & {} &  
$f(T)^{\sum_{I=1}^{s}k_I^2}$ &
{} \\ \hline\hline
$\mathcal{C}$ &
$\left( \prod_m 2\pi \right)^{\frac{9}{2}N^2}$ & 
$f(T)^{-8N^2} \cdot g(T)^{8N^2}$ &
$\textrm{Vol}(U(N))_{\textrm{local}}$ \\ \hline
\end{tabular}
\caption{\footnotesize 
The list of contributions to the overall factor 
of the partition function for general classical vacua.}
\label{factor2}
}
The overall factor of the partition function $\mathcal{C}$ 
turns out to the same for all the classical vacua, and it 
is given by
\begin{eqnarray} 
\mathcal{C} \equiv 
	\left(\prod_{m} 2 \pi\right)^{\frac{9}{2}N^2} 
	\cdot f(T)^{-8N^2} \cdot g(T)^{8N^2} 
	\cdot \textrm{Vol}(U(N))_{\textrm{local}} \ .
\end{eqnarray} 

\section{Evaluating free energy with the eigenvalue density}
\label{section:eval-free}

In this section we derive the free energy (\ref{free-energy_exact})
for the eigenvalue density (\ref{eigenvalue-distribution_exact}).
Omitting the $m\ge 2$ terms in (\ref{effective-action4}),
the effective action is given as
\bear
S_{\rm eff} &\approx& 
- N^2 \int_{-\theta_{\rm cl}}^{\theta_{\rm cl}} 
\int_{-\theta_{\rm cl}}^{\theta_{\rm cl}} 
d\theta d\theta' \rho(\theta) \rho(\theta') 
f_1(\beta)\cos{(\theta-\theta')} \nn \\
&& -N^2 {\cal P.V.} \int_{-\theta_{\rm cl}}^{\theta_{\rm cl}} 
\int_{-\theta_{\rm cl}}^{\theta_{\rm cl}} 
d\theta d\theta' \rho(\theta) \rho(\theta') 
\ln \left| \sin\left(\half (\theta-\theta')\right) \right| \nn \\
&& +N^2 \frac{1}{2\pi} \int_{-\pi}^{\pi} d\theta 
\ln\left| \sin \frac{\theta}{2} \right|\, ,
\eear
where the symbol ${\cal P.V.}$ represents the principal value.
For eigenvalue density (\ref{eigenvalue-distribution_exact}) 
one can easily obtain
\bear
&& 
\int_{-\theta_{\rm c}}^{\theta_{\rm cl}} d\theta \rho(\theta) \sin\theta 
= 0 \ , \quad
\int_{-\theta_{\rm c}}^{\theta_{\rm cl}} d\theta \rho(\theta) \cos\theta 
= 1-\frac{\omega}{4} \ , \quad
\int_{-\pi}^{\pi}d\theta\ln \left| \sin \frac{\theta}{2} \right| 
= - \ln 2 \ , \\
&& {\cal P.V.} \int_{-\theta_{\rm cl}}^{\theta_{\rm cl}} 
d\theta d\theta'\rho(\theta)\rho(\theta')
\ln \left|  \sin 
\frac{\theta-\theta'}{2}
 \right| 
= -\ln 2 +\half \ln\frac{\omega}{2} - \quarter 
\label{int-rho4}
\, ,
\eear
where, in deriving the formula in the second line, 
we have used the fact that the distribution
(\ref{eigenvalue-distribution_exact}) satisfies
the saddle-point equation (\ref{sad-point-eq}).
Using these formulae, we obtain 
\be
-\frac{S_{\rm eff}}{N^2} \approx f_1(\beta) \left(1-\frac{\omega}{4}\right)^2 
+\half\ln\frac{\omega}{2} - \quarter 
= \frac{f_1(\beta)}{2}\left(2-\frac{\omega}{2}\right)
+\half\ln\frac{\omega}{2} - \half \ .
\ee

\end{document}